\documentclass[11pt, 5p, nopreprintline]{elsarticle}

\journal{TBA}

\biboptions{numbers,sort&compress}

\usepackage{libertine}
\usepackage{libertinust1math}
\usepackage{amsmath}                 
\usepackage{bbold}                   
\usepackage{graphicx}                
\usepackage{eurosym}                 
\usepackage{mathtools}               
\usepackage{url}                     
\usepackage{booktabs}                
\usepackage{epstopdf}                
\usepackage{xfrac}                   
\usepackage{tabularx}                
\usepackage{bm}                      
\usepackage{subcaption}              
\usepackage{longtable}               
\usepackage{multirow}                
\usepackage{threeparttable}          
\usepackage{pdflscape}               
\usepackage[svgnames]{xcolor}    	 
\usepackage[export]{adjustbox}       
\usepackage[version=4]{mhchem}       
\usepackage{xurl} 				     
\usepackage[colorlinks]{hyperref}    
\usepackage[parfill]{parskip}        
\usepackage[nameinlink,sort&compress,capitalise,noabbrev]{cleveref} 
\usepackage[leftcaption,raggedright]{sidecap} 
\usepackage[prependcaption,textsize=footnotesize]{todonotes} 
\usepackage{siunitx}                 
\usepackage{csquotes} 				 
\usepackage[acronym, automake, nonumberlist]{glossaries-extra}
\usepackage{svg} 				    
\usepackage{placeins} 			  	
\usepackage{multicol}			    
\usepackage{stfloats}

\usepackage{float}
\usepackage{lipsum}
\usepackage{mdframed}
\usepackage[resetlabels,labeled]{multibib} 
\newcites{S}{Supplementary References}
\bibliographystyleS{elsarticle-num}

\DeclareSIUnit\year{a}
\DeclareSIUnit{\tco}{t_{\ce{CO2}}}
\DeclareSIUnit{\sieuro}{\mbox{\euro}}

\graphicspath{ 						
	{../results/paper-wp1},
	}

\newcommand{\co}{\ce{CO2}}

\newcommand{\ubar}[1]{\text{\b{$#1$}}} 

\newdefinition{rmk}{Remark}

\definecolor{darkred}{RGB}{156, 0, 0}

\makeglossaries

\newacronym{eu}{EU}{European Union}
\newacronym{bdew}{BDEW}{Bundesverband der Energie- und Wasserwirtschaft}
\newacronym{dea}{DEA}{Danish Energy Agency}
\newacronym{jrc}{JRC}{Joint Research Center}
\newacronym{tyndp}{TYNDP}{Ten-Year Network Development Plan}
\newacronym{entsoe}{ENTSO-E}{European Network for Transmission System Operators for Electricity}
\newacronym{entsog}{ENTSO-G}{European Network of Transmission System Operators for Gas}

\newacronym{tes}{TES}{Thermal Energy Storage (in form of water tanks)}
\newacronym{chp}{CHP}{Combined Heat and Power plant}
\newacronym{cop}{COP}{Coefficient of Performance}
\newacronym{PV}{PV}{Solar photovoltaics}
\newacronym{ghg}{GHG}{Greenhouse gas}
\newacronym{helmeth}{HELMETH}{Integrated High-Temperature Electrolysis and Methanation for Effective Power to Gas Conversion}
\newacronym{ocgt}{OCGT}{Open Cycle Gas Turbine}
\newacronym{v2g}{V2G}{Vehicle to Grid}
\newacronym{dsm}{DSM}{Demand Side Management}
\newacronym{hvdc}{HVDC}{High Voltage Direct Current}
\newacronym{phs}{PHS}{Pumped Hydro Storage}
\newacronym{dac}{DAC}{Direct Air Capture}
\newacronym{sng}{SNG}{Synthetic Natural Gas}
\newacronym{res}{RES}{Renewable Energy Sources}
\newacronym{cfe}{CFE}{Carbon-Free Electricity}
\newacronym{ldes}{LDES}{Long Duration Energy Storage}
\newacronym{ccs}{CCS}{Carbon Capture and Sequestration}

\newacronym{fom}{FOM}{Fixed Operation and Maintenance costs}
\newacronym{vom}{VOM}{Variable Operation and Maintenance costs}
\newacronym{capex}{CAPEX}{Capital expenditures}
\newacronym{opex}{OPEX}{Operational expenditures}
\newacronym{ppa}{PPA}{Power Purchase Agreement}
\newacronym{ptc}{PTC}{Production Tax Credit}
\newacronym{necp}{NECP}{National Energy and Climate Plan}
\newacronym{ets}{ETS}{Emissions Trading System}
\newacronym{cfd}{CfD}{Contract for Difference}
\newacronym{go}{GO}{Guarantees of Origin}
\newacronym{rec}{REC}{Renewable Energy Certificate}
\newacronym{lcoe}{LCOE}{Levelised Cost of Electricity}

\newacronym{pypsa}{PyPSA}{Python for Power System Analysis}
\newacronym{nuts}{NUTS}{Nomenclature of Territorial Units for Statistics}
\newacronym{ci}{C\&I}{Corporate and Industry (sectors)}
\newacronym{ngo}{NGO}{Non-Governmental Organization}
\newacronym{ict}{ICT}{Information and Communications Technology}

\begin{document}

\begin{frontmatter}

	\title{Spatio-temporal load shifting for truly clean computing}

	\author[tubaddress]{Iegor Riepin\fnref{fn1}}
	\ead{iegor.riepin@tu-berlin.de}
	\author[tubaddress]{Tom Brown}
	\ead{t.brown@tu-berlin.de}
	\author[wiaddress]{Victor M. Zavala}
	\ead{victor.zavala@wisc.edu}

	\address[tubaddress]{Department of Digital Transformation in Energy Systems, TU Berlin, Germany}
	\address[wiaddress]{%
  		\begin{tabular}[t]{@{}l@{}}
			Department of Chemical and Biological Engineering, University of Wisconsin-Madison, USA \\
			\quad\quad Mathematics and Computer Science Division, Argonne National Laboratory, USA
		\end{tabular}%
		}
    \fntext[fn1]{Corresponding author: Iegor Riepin, TU Berlin, Einsteinufer 25, 10587 Berlin, Germany}

	\begin{abstract}
%

Companies with datacenters are procuring significant amounts of renewable energy to reduce their carbon footprint.
There is increasing interest in achieving 24/7 Carbon-Free Energy (CFE) matching in electricity usage, aiming to eliminate all carbon footprints associated with electricity consumption on an hourly basis.
However, the variability of renewable energy resources poses significant challenges for achieving this goal.
We explore the impact of shifting computing jobs and associated power loads both in time and between datacenter locations.
We develop an optimization model to simulate a network of geographically distributed datacenters managed by a company leveraging spatio-temporal load flexibility to achieve 24/7 CFE matching.
We isolate three signals relevant for informed use of load flexiblity: varying average quality of renewable energy resources, low correlation between wind power generation over long distances due to different weather conditions, and lags in solar radiation peak due to Earth's rotation.
We illustrate that the location of datacenters and the time of year affect which signal drives an effective load-shaping strategy.
The energy procurement and load-shifting decisions based on informed use of these signals facilitate the resource-efficiency and cost-effectiveness of clean computing---the costs of 24/7 CFE are reduced by 1.29$\pm$0.07~\euro/MWh for every additional percentage of flexible load.
We provide practical guidelines on how companies with datacenters can leverage spatio-temporal load flexibility for truly clean computing.
Our results and the open-source optimization model can also be useful for a broader variety of companies with flexible loads and an interest in eliminating their carbon footprint.
	\end{abstract}



\end{frontmatter}

\tableofcontents

\section{Introduction}
\label{sec:intro}
%



Demand for digital services is growing rapidly. The number of internet users has doubled worldwide over the past decade, while traffic has grown 25-fold \cite{ieaDataCentresData2023}.
The significant energy consumption of Information and Communications Technology (\gls{ict}) infrastructure is becoming a concern from an ecological standpoint.
Global datacenter energy use is estimated to be between 1\% and 2\% of final electricity demand worldwide in 2020 \cite{davidmyttonHowMuchEnergy2020, masanetRecalibratingGlobalData2020} and is likely to increase rapidly in the future \cite{andraeGlobalElectricityUsage2015}.
The Greenhouse Gas (\gls{ghg}) emissions footprint of the datacenters and data transmission networks was estimated to be 330~million tones (Mt) \co-eq in 2020, equivalent to 0.9\% of global energy-related emissions \cite{ieaDataCentresData2023, malmodinICTSectorElectricity2023}.
For perspective, the national CO$_2$ emissions of the United Kingdom were around 331~Mt in 2021 \cite{UKnationalstats}.


The growing demand for computing drives innovation and economies of scale. Large companies, such as Amazon, Google, and Microsoft are centralizing computing infrastructure into \enquote{hyperscale} datacenters.
Such large and highly efficient facilities consist of hundreds of server nodes that are managed collectively and geographically dispersed throughout the world \cite{ThereAre500}.

In their role as electricity consumers, hyperscale datacenters have unique characteristics when it comes to demand-side management.
Some computing jobs and associated power loads are \enquote{flexible}, i.e., the jobs are not time-critical and can be scheduled flexibly without impacting the overall quality of service.
Therefore, datacenter operators have the ability to shift a considerable fraction of computing workloads \textit{in~time} and \textit{in space} \cite{radovanovicIEEE2023}.
The temporal load shifting implies re-scheduling of flexible workloads to another time point, i.e., delaying computing job execution.
The spatial load shifting implies migration of flexible compute jobs and associated power loads between different physical datacenter locations.

The significant energy consumption of \gls{ict} infrastructure, and its anticipated growth, raise a natural question: \textit{How can datacenters make use of their unique spatial and temporal load-shifting capabilities to improve their economics and the environmental footprint?} The topic of this overarching question has been the subject of a growing body of literature over the past decade.

\textbf{Existing literature --} The work by Wierman et al.  \cite{wiermanOpportunitiesChallengesData2014} surveyed the forms, opportunities and challenges for data center demand response options, including aspects of temporal load shifting such as workload delaying or shedding and the geographical load balancing. Several papers discussed methods and policies needed to incentivise parcicipation of datacenters in the power grid's demand responce \cite{liuPricingDataCenter2014, zhouTruthfulEfficientIncentive2020}.

Numerous studies investigated ways to harness spatial or temporal load-shifting flexibility to achieve certain economical or environmental goals. Many studies have proposed workload scheduling algorithms for geographical load shifting to reduce datacenter electricity costs \cite{velascoElasticOperationsFederated2014, douCarbonAwareElectricityCost2017, heMinimizingOperationCost2021, raoDistributedCoordinationInternet2012, renCarbonAwareEnergyCapacity2012, dengEcoAwareOnlinePower2016}, or looked at the potential of datacenter load migration for mitigating renewable curtailment and \gls{ghg} emissions in the local electricity grid \cite{zhengMitigatingCurtailmentCarbon2020, mahmudDistributedFrameworkCarbon2016}, or aimed to increase the share of renewable energy in datacenter electricity consumption \cite{wangGreenawareVirtualMachine2015, kimDataCentersDispatchable2017, liuGeographicalLoadBalancing2011, kellyBalancingPowerSystems2016}.

Recently, several papers elaborated a mathematical problem that captures both spatial and temporal load-shifting flexibility. Zhang et al. \cite{zhangFlexibilityNetworksData2020} introduced the concept of virtual links to capture space-time load flexibility provided by geographically-distributed data centers in market clearing procedures. This work was followed by a paper in which the authors demonstrated a market clearing formulation that seeks to remunerate spatio-temporal load shifting for the flexibility service datacenters can offer to the power grid \cite{zhangRemuneratingSpaceTime2022}.

A great deal of attention in previous research exploring the signals for geographical load shifting has focused on information characterising the state of electricity grids where datacenters operate. These signals include average carbon emissions in the region of operation \cite{zhengMitigatingCurtailmentCarbon2020}, or locational marginal carbon emissions \cite{lindbergEnvironmentalPotentialHyperScale2021}, or locational electricity prices and price differences across datacenter locations \cite{raoMinimizingElectricityCost2010,tranHowGeoDistributedData2016, zhangRemuneratingSpaceTime2022}. This emerging stream of research work is now finding practical applications. For example, Google recently introduced a Carbon-Intelligent Compute Management system that minimizes carbon footprint and power infrastructure costs by shifting flexible workloads in time  across datacenter fleet as a function to the next day's carbon intensity forecasts \cite{radovanovicIEEE2023}. There are several companies providing market data and forecasts on carbon emissions and electricity prices, such as WattTime \cite{WattTime}, ElectricityMaps \cite{ElectricityMaps}, and others.

\textbf{24/7 Carbon-Free Energy matching --} \gls{ict} companies invest significantly in renewable energy to reduce their environmental impact, reduce power price volatility, and enhance their brand image. The hyperscale datacenter industry in particular leads in corporate renewable energy acquisition through Power Purchase Agreements (\gls{ppa}s). With almost 50~GW contracted so far, Amazon, Microsoft, Meta, and Google are the four largest purchasers of corporate renewable energy \gls{ppa}s \cite{ieaDataCentresData2023}. Furthermore, some companies have pledged to eliminate \textit{all} greenhouse gas emissions associated with their electricity use. These companies have set a \enquote{24/7 Carbon-Free Energy (\gls{cfe})} goal aiming to match electricity demand with clean energy supply on an \textit{hourly basis}. The 24/7 CFE targets were announced by Google and Microsoft for 2030 and Iron Mountain for 2040 \cite{google-247by2030, Microsoft-vision, IronMountainSustainability}.

One notable feature of the 24/7 CFE goal is that it drives---and requires---a significant volume of energy procurement via \gls{ppa}s. Recent research showed that achieving 24/7 CFE with a 100\% target (i.e., achieving 0~gCO$_2$ emissions per kWh) implies that companies need to rely on their own procured resources,without any electricity procurement from the local grid \cite{riepinMeansCostsSystemlevel2023, riepin-zenodo-systemlevel247}. This is because the local electricity mix in most regions and at most times does not have a strictly zero carbon content and therefore cannot contribute to 24/7 CFE.

\textbf{The problem --}
Companies aiming for 24/7 clean electricity have a unique challenge regarding load-shifting. Because they rely on their own portfolio of carbon-free energy generators instead of buying electricity from a local grid, grid signals such as average carbon emissions or locational electricity prices no longer play a significant role for informed load-shifting. This effect was illustrated with the help of energy system modelling in the previous work of the authours \cite{riepin-zenodo-systemlevel247,riepinMeansCostsSystemlevel2023}. In this context, the following question arise: \textit{In pursuit of achieving 24/7 Carbon-Free Energy objectives, what signals should datacenter operators focus on to facilitate informed and effective load shifting across space and time?}

\textbf{Contribution --} This is the first paper to examine the role of space-time load shifting in the context of 24/7 carbon-free energy matching.
Through energy system modelling, we identify signals for effective shifting of load across space and time. We show that taking these signals into consideration in load shifting and energy procurement strategies can enable companies to achieve significant gains in energy efficiency and affordability of truly clean computing.

For that, we develop a computer model that simulates a datacenter fleet with controlled degrees of flexibility in pursuit of 24/7 CFE objectives. Datacenter fleet optimization is incorporated into a European electricity system model as a mathematical problem.
We identify three individual signals that are relevant for effective geographical and temporal load shifting. These signals include (a) varying quality of renewable energy resources (i.e. average capacity factors) across datacenter locations, (b) low correlation between wind power generation over long distances due to different weather conditions, and (c) lags in solar radiation peak due to Earth's rotation. The results show that each of these signals plays a role in effective load shifting, although the \enquote{weight} of each signal in optimal load shaping strategies depends on the locations of datacenters and time of the year.
Furthermore, we demonstrate that load shifting decisions optimized considering these signals facilitate efficiency and affordability of the 24/7 CFE matching. When all three signals are taken into account, the space-time load shifting can reduce the electricity costs of a datacenter fleet operating at 0~kgCO$_2$/MWh by 1.29$\pm$0.07~\euro/MWh for each percent point of flexible load.


The remainder of the paper is structured as follows: \cref{sec:methods} introduces a mathematical model of clean energy procurement and gives a brief summary of the methodology, sources of model input data, and the experimental setup. \cref{sec:results} presents the results of the experiments and discusses the implications of the findings. \cref{sec:discussion} discusses the results in the context of the existing literature, practical recommendations for incorporating discussed signals into load shaping strategies and gives critical appraisal of the results. \cref{sec:conclusion} concludes the paper. All code necessary to reproduce the experiments is published under an open license alongside the paper. See \nameref{sec:code}.

\section{Methods and experimental design}
\label{sec:methods}
%

\subsection{Methods and tools}

The simulations were carried out with the open-source software PyPSA for energy system modelling \cite{horschPyPSAEurOpenOptimisation2018} and Linopy for optimization \cite{LinopyLinearOptimization2024}.
The mathematical model is a system-wide cost-minimisation problem.
The objective of the model is to co-optimise (i) investment and dispatch decisions of generation and storage assets done by datacenters to meet their electricity demand in line with 24/7 CFE objectives, (ii) space-time load-shifting decisions subject to datacenter flexibility constraints, as well as (iii) investment and dispatch decisions of assets in the rest of the European electricity system to meet the demand of other consumers.
The model formulation includes the linear optimal power flow approximation on the transmission network.
This work uses a brownfield investment approach, which means that the model includes information about the existing assets of the European electricity system.

The electricity system dispatch and investment problem of this type is a standard in the energy modelling literature \cite{OpenModelsWikib}. The clean computing model is based on the framework of 24/7 CFE accounting suggested by Google \cite{google-methodologies} that was implemented in realms of energy system models by Xu et al. \cite{xu-247CFE-report} and Riepin \& Brown \cite{riepinMeansCostsSystemlevel2023}. The space-time load flexibility model is inspired by the work of Zhang et al. \cite{zhangRemuneratingSpaceTime2022}. Finally, the space-time load flexibility as a degree of freedom within the 24/7 CFE matching problem is a novel contribution of this work.

\subsection{Datacenter model}

Datacenters are represented as a subset of demand nodes for electricity with a fixed demand profile, controlled degree of flexibility, and a set of specific constraints ensuring the carbon-free energy matching and utilisation of flexibility in a feasible domain.

\begin{figure}[b]
    \centering
    \includegraphics[width=1\columnwidth]{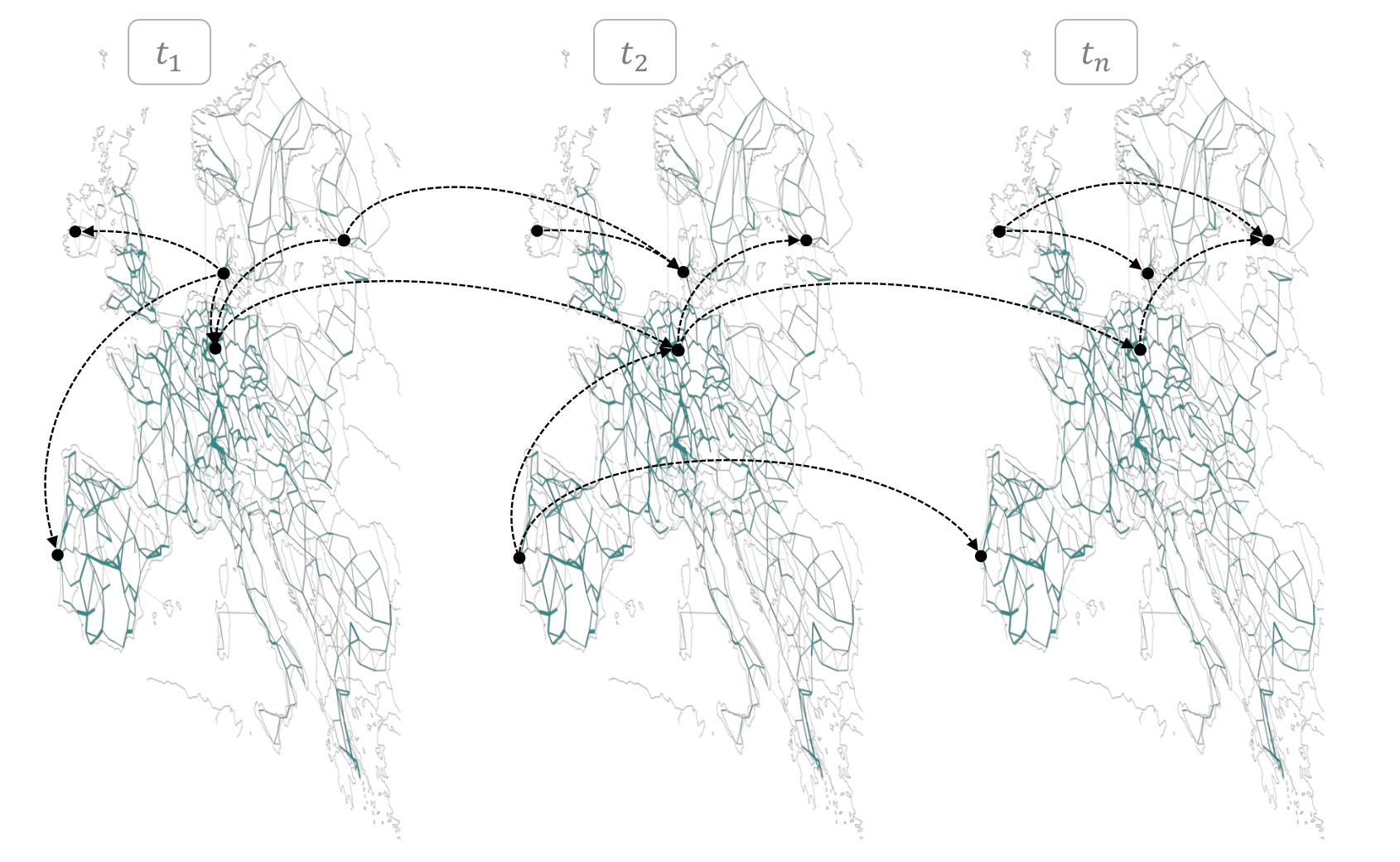}
    \caption{Illustration of spatio-temporal load-shifting optimisation problem faced by a datacenter operator.}
    \label{fig:space-time-optimisation}
\end{figure}

The 24/7 CFE matching problem introduces a set of new model components (parameters, variables, constraints) into the power system optimisation problem to represent voluntary yet binding carbon-free energy matching commitments. Here, we model a situation where one company operates a network of datacenters that are geographically distributed but managed collectively. The company is committed \textit{to match every kilowatt-hour of electricity consumption by carbon-free sources in all datacenter locations}. To achieve the target, the company optimises its procurement of carbon-free generation and storage resources, dispatch of procured assets, imports of electricity from the regional grid, spatio-temporal load-shifting flexibility use (schematically illustrated in Figure \ref{fig:space-time-optimisation}), and when necessary, curtailment of excess generation.

\textbf{A single datacenter with an inflexible load --} First, consider a case of a single datacenter with demand profile $d_{t}$ for each hour $t$ of the year. The demand profile is assumed to be known in advance and is not flexible. In such case, energy balance constraint requires that hourly demand is met by a combination of the following: (i) dispatch $g_{r,t}$ of procured carbon-free generators $r\in CFE$, (ii) dispatch $\bar{g}_{s,t}$ of procured storage technologies $s\in STO$, and (iii) imports of electricity from the regional grid $im_{t}$:
\begin{equation}
    \begin{split}
        \sum_{r\in CFE} g_{r,t} &+ \sum_{s\in STO} \left(\bar{g}_{s,t} - \underline{g}_{s,t}\right) - ex_t + im_t = d_t \quad \forall t \in T
    \end{split}
\label{eqn:inflexnb}
\end{equation}
Note that if total electricity yield of renewable generators procured by the datacenter operator exceeds demand in a given hour, the excess electricity $ex_t$ has to be either stored or curtailed.\footnote{In practice, excess electricity can also be sold to the regional electricity market at wholesale market prices. This option is deliberately avoided in this work. When included, optimal usage of flexibility factors in potential revenues from selling excess electricity to the grid.}

The 24/7 CFE matching constraint requires that sum over hourly generation from the contracted generators, plus net dispatch of the storage technologies, plus imports of electricity from the regional grid multiplied by the grid's hourly CFE factor $CFE_t$ minus the excess electricity must be higher or equal than a certain CFE score $x$ multiplied by the total load of a datacenter:
\begin{equation}
    \begin{split}
        \sum_{r\in CFE, t\in T} g_{r,t} &+ \sum_{s\in STO, t\in T} \left(\bar{g}_{s,t} - \underline{g}_{s,t}\right) \\
        &- \sum_{t\in T} ex_t + \sum_{t\in T} CFE_t \cdot im_t \geq x \cdot \sum_{t\in T} d_t
    \end{split}
\label{eqn:CFE}
\end{equation}
\noindent where \textit{CFE score} $x$[\%] measures the degree to which hourly electricity consumption is matched with carbon-free electricity generation \cite{google-methodologies}. Thus, equation (\ref{eqn:CFE}) allows for controlling \textit{the quality score} of the 24/7~CFE procurement by adjusting the parameter $x$, which was the subject of research in recent literature \cite{riepinMeansCostsSystemlevel2023,xu-247CFE-report}. Here, we focus on the best quality score ($x=1$) ensuring that every kilowatt-hour of electricity consumption is met by carbon-free sources at all times.

Note the following properties of the 24/7 CFE matching problem:

\begin{enumerate}
    \item The contracted generators $r\in CFE$ must be additional to the system and can be sited only in the local bidding zone, known as the requirements for \textit{additionality} and \textit{locational matching}.
    \item Excess electricity is not counted toward the CFE score and thus subtracted from the left-hand side of eq. (\ref{eqn:CFE}).
    \item The CFE factor of the regional grid ($CFE_t$) can be seen as the percentage of clean electricity in each MWh of imported electricity to supply demand of participating consumers in a given hour.
    To compute $CFE_t$, we consider both the hourly electricity mix in the local bidding zone and emission intensity of imported electricity from the neighbouring zones.
    The numerical simulations show that for the perfect 24/7 CFE matching ($x=1$), the datacenter operator does not rely on electricity imports from the regional grid because local electricity mix must have a strictly zero carbon content for imported electricity to be counted as \enquote{carbon-free}.
    Thus, the datacenter operator has one less degree of freedom to meet the 24/7 CFE matching target.
    The methodology to calculate the grid CFE factor and linearise resulting nonconvexity is described in detail in the prior work of the authors \cite{riepin-zenodo-systemlevel247}.
\end{enumerate}

\textbf{Multiple datacenters with flexible load --} Here we expand the 24/7 CFE matching problem to a case of multiple datacenters with flexible loads.

\begin{figure}
    \centering
    \includegraphics[width=1\columnwidth]{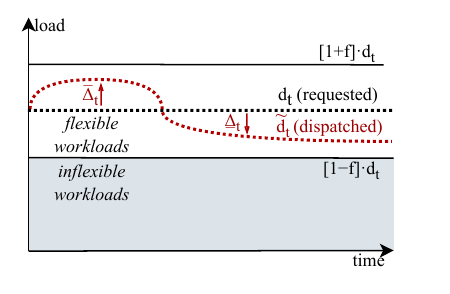}
    \caption{Illustration of load flexibility concept: the absolute value of deviations between dispatched $\widetilde{d}_t$~[MW] and requested $d_t$~[MW] loads must fall within a certain flexibility range $f$~[\%] for each hour $t$.}
    \label{fig:workloads}
\end{figure}

Let us assume that a known number of computing jobs with their associated power demand are \enquote{flexible}, i.e., power demand can potentially be shifted geographically across datacenter locations, or delayed to other times.\footnote{The amount of CPU usage on a cluster can be mapped accurately to electricity demand \cite{radovanovicIEEE2023}.} Thus, the \textit{dispatched load} $\widetilde{d}_t$ of a datacenter can deviate from the requested load $d_t$. The dispatched load $\widetilde{d}_t$ can take a value in a feasible domain that is constrained by a datacenter capacity (an upper limit) and the inflexible workloads volume (a lower limit), as illustrated in Figure \ref{fig:workloads}. The range of possible deviations between the dispatched and the requested loads is assumed to lie within a certain \textit{flexibility range} $f$~[\%], such as:
\begin{subequations}
  \begin{align}
    [1-f] \cdot d_t \le  \widetilde{d}_t  \le [1+f] \cdot d_t \quad \forall t \in T
    \label{eqn:dcaps} \\
    \widetilde{d}_t = d_t + (\overline{\Delta}_t - \underline{\Delta}_t) \quad \forall t \in T
    \label{eqn:dtilde}
  \end{align}
  \label{eqn:range}
\end{subequations}
\noindent where $\overline{\Delta}_t, \underline{\Delta}_t \in \mathbb{R}_{+}$ stand for positive/negative deviation of $\widetilde{d}_t$ and $d_t$ in hour $t$.

To incorporate spatial flexibility, we use a concept of \textit{virtual links} representing non-physical pathways for shifting power loads across datacenter locations. Assume we have a set of datacenters $N$ located in various locations of an electricity network. Let $\Theta$ be the set of all virtual links and let $\delta_\vartheta \in \mathbb{R}_{+}$ be the spatial load shifts (flows via virtual pathways). We can thus define $\Theta_n^{snd} := \{\vartheta \in \Theta | snd(\vartheta) = n\} \subseteq \Theta$ and $\Theta_n^{rec} := \{\vartheta \in \Theta | rec(\vartheta) = n\} \subseteq \Theta$ to be the set of sending and receiving virtual links for each datacenter $n \in N$.


To incorporate temporal flexibility, we use a concept of \textit{temporal load management} that allows for shifting load from one time point to another in the future.\footnote{Zhang et al. \cite{zhangRemuneratingSpaceTime2022} show that the temporal load shifting problem can also be formulated via the concept of \textit{virtual links}. This yields the same mathematical problem.} Let each datacenter have a temporal load management mechanism $S_n^{dsm} := \{s' \in S | dsm(s') = n\}$. If $T = \{t_1 , t_2 , ..., t_T\}$ is a time horizon of our optimization problem, we can define $\bar{g}_{s',n,t}, \ubar{g}_{s',n,t} \in \mathbb{R}_{+}$ to be the amount of load shifted from time $t$ to another time point $t'$ for each datacenter $n \in N$ and $t, t' \in T$.


The nodal energy balance for inflexible consumers (eq.~\ref{eqn:inflexnb}) can now be extended by variables representing shifts of load \textit{in space} and \textit{in time}. Thus, the dispatched load  $\widetilde{d_{t}}$ of flexible consumer can deviate from the $d_{n,t}$ value due to space-time load shifting.
\begin{equation}
    \begin{split}
    &\sum_{r\in CFE} g_{r,n,t} + \sum_{s\in STO} \left(\bar{g}_{s,n,t} - \ubar{g}_{s,n,t}\right) - ex_{n,t} + im_{n,t}  = \\
    & d_{n,t} + \textcolor{darkred}{\sum_{\vartheta \in \Theta_n^{rec}}\delta_{\vartheta, t} - \sum_{\vartheta \in \Theta_n^{snd}}\delta_{\vartheta, t} + \sum_{{s'} \in S_n^{dsm}} \left(\bar{g}_{s',n,t} - \ubar{g}_{s',n,t}\right)} \\
    & \hspace{.5cm} \forall n \in N, t \in T
    \label{eqn:bothnb}
    \end{split}
\end{equation}
Computing capacity constraints (eq.~\ref{eqn:bothflex}) ensure that the dispatched load at each datacenter  $\widetilde{d}_{n,t}$ falls within the feasible domain considering both spatial and temporal load shifting. Computing capacity limit sets an upper bound (eq. \ref{eqn:bothb}) and inflexible workloads share sets a lower bound (eq. \ref{eqn:bothc}).
\begin{subequations}
    \begin{align}
      \begin{split}
        &\widetilde{d}_{n,t} =  d_{n,t} + \sum_{\vartheta \in \Theta_n^{rec}}\delta_{\vartheta, t} - \sum_{\vartheta \in \Theta_n^{snd}}\delta_{\vartheta, t} \\
        &+ \sum_{{s'} \in S_n^{dsm}} \left(\bar{g}_{s',n,t} - \ubar{g}_{s',n,t}\right) \quad \forall n \in N, t \in T \\
      \end{split}
      \label{eqn:botha} \\
      &\widetilde{d}_{n,t} \le [1+f] \cdot d_{n,t}  \quad \forall n \in N, t \in T \label{eqn:bothb} \\
      &\widetilde{d}_{n,t} \ge [1-f] \cdot d_{n,t}  \quad \forall n \in N, t \in T \label{eqn:bothc}
    \end{align}
    \label{eqn:bothflex}
\end{subequations}
Note that spatial load shifts are not subject to any electricity network transmission constraints. As such, the only source of congestion for the virtual links is computing capacity constraints (i.e., availability of flexible workloads) as defined in eq. (\ref{eqn:bothflex}).

Temporal load shifts have an additional constraint ensuring that the cluster-level compute usage is preserved within a certain time interval. We follow the work of Radovanovic et al. \cite{radovanovicIEEE2023} and implement a \textit{daily} compute usage conservation rule:
\begin{equation}
    \sum_{t | t \in t(DAYS)} \left(\bar{g}_{s',t} - \ubar{g}_{s',t}\right) = 0 \quad \forall {s'} \in S_n^{dsm}
    \label{eqn:dailyconserv2}
\end{equation}
Finally, the 24/7~CFE matching constraint defined in eq. \ref{eqn:CFE} can be defined over a set of datacenters nodes $N$ and extended on the right-hand side with variables representing spatial and temporal load shifts:
\begin{align}
    &\sum_{r\in CFE, t\in T} g_{r,n,t} + \sum_{s\in STO, t\in T} (\bar{g}_{s,n,t} - \ubar{g}_{s,n,t}) \nonumber \\
    &- \sum_{t\in T} ex_{n,t} + \sum_{t\in T} CFE_{n,t} \cdot im_{n,t} \geq \nonumber \\
    &x_n \cdot
        \Bigg( \sum_{t\in T} \Big( d_{n,t} + \textcolor{darkred}{\sum_{\vartheta \in \Theta_n^{rec}}\delta_{\vartheta, t} - \sum_{\vartheta \in \Theta_n^{snd}}\delta_{\vartheta, t}} \nonumber \\
        & \textcolor{darkred}{+ \sum_{{s'} \in S_n^{dsm}} (\bar{g}_{s',n,t} - \ubar{g}_{s',n,t})}\Big)\Bigg) \quad \forall n \in N\label{eqn:bothCFE}
\end{align}
Overall, the mathematical formulation of the 24/7 CFE matching problem extended with eqs. \ref{eqn:bothnb}-\ref{eqn:bothCFE} represents a situation where a company pursuing the 24/7 CFE matching target has two additional degrees of freedom---the ability to shift power loads in space and in time---as part of its decision-making process. Co-optimisation of space-time load-shifting flexibility with procurement and dispatch of carbon-free generators can lead to a more resource-efficient and cost-effective solution for carbon-free electricity matching.

\subsection{Model scope and parametrisation}

Geographical focus of the model covers the entire European electricity system clustered to 37 zones. Each zone represents an individual country. Some countries that straddle different synchronous areas are split to individual bidding zones, such as DK1 (West) and DK2 (East). A set of datacenters---electricity consumers committed to 24/7 CFE goals---are placed in several selected zones in each scenario. Each zone have unique weather patterns, renewable potentials, and background electricity mixes. The model's time horizon spans one year. All model runs are done with 1-hourly temporal resolution, i.e., 8760 time steps.

We assume all datacenters to have a load of 100~MW and a flat (baseload) consumption profile. Note that the shape of a consumption profile does not significantly impact the costs of the 24/7 CFE matching \cite{riepin-zenodo-systemlevel247}. We assume a \textit{complete graph} superstructure of virtual links. Specifically, for every pair of distinct datacenters \(n, m \in N\) where \(n \neq m\), there exists a bidirectional virtual link \(\vartheta_{nm} \in \Theta\) such that \(snd(\vartheta_{nm}) = n\) and \(rec(\vartheta_{nm}) = m\), and conversely, \(snd(\vartheta_{mn}) = m\) and \(rec(\vartheta_{mn}) = n\). This configuration ensures that each datacenter is capable of both sending and receiving spatial load shifts to and from every other datacenter in the network.

The model is parametrised for the exemplary year 2025. Input data such as technology cost assumptions, national renewable policies, decommissioning of legacy power plant fleet, and system-wide assumptions are parametrised accordingly. For the 24/7 CFE matching problem, we assume a set of energy technologies that \textit{are commercially available today}: onshore wind, utility scale solar photovoltaics (\gls{PV}), and battery storage. Cost and other assumptions for these technologies are collected from the Danish Energy Agency \cite{DEA-technologydata} and provided in \nameref{sec:si}.

Weather-dependent power potentials for renewable energy technologies, including those of solar PV and onshore wind generators available for the 24/7 CFE matching, are simulated with hourly temporal resolution using the ERA5 reanalysis dataset \cite{hersbachERA5GlobalReanalysis2020} using the open-source tool Atlite \cite{atlite-github}. We assume 2013 to be a representative climate year.

Other electricity system data is processed with an open-source model for the European energy system PyPSA-Eur \cite{PyPSAEur-docs}. All code and input data to reproduce the experiments are available at the GitHub repository under open licenses \cite{github-spacetime}.

\section{Results}
\label{sec:results}
%

We present the results as a sequence of case studies of varying levels of complexity, since there are several interacting signals at work. In the first three subsections, we isolate three particular signals that affect the load-shifting. In the final section, we generalize the results across any combination of datacenters.

\subsection{Signal 1: quality of local renewable resources}

\begin{figure*}
    \centering
    \includegraphics[width=\textwidth]{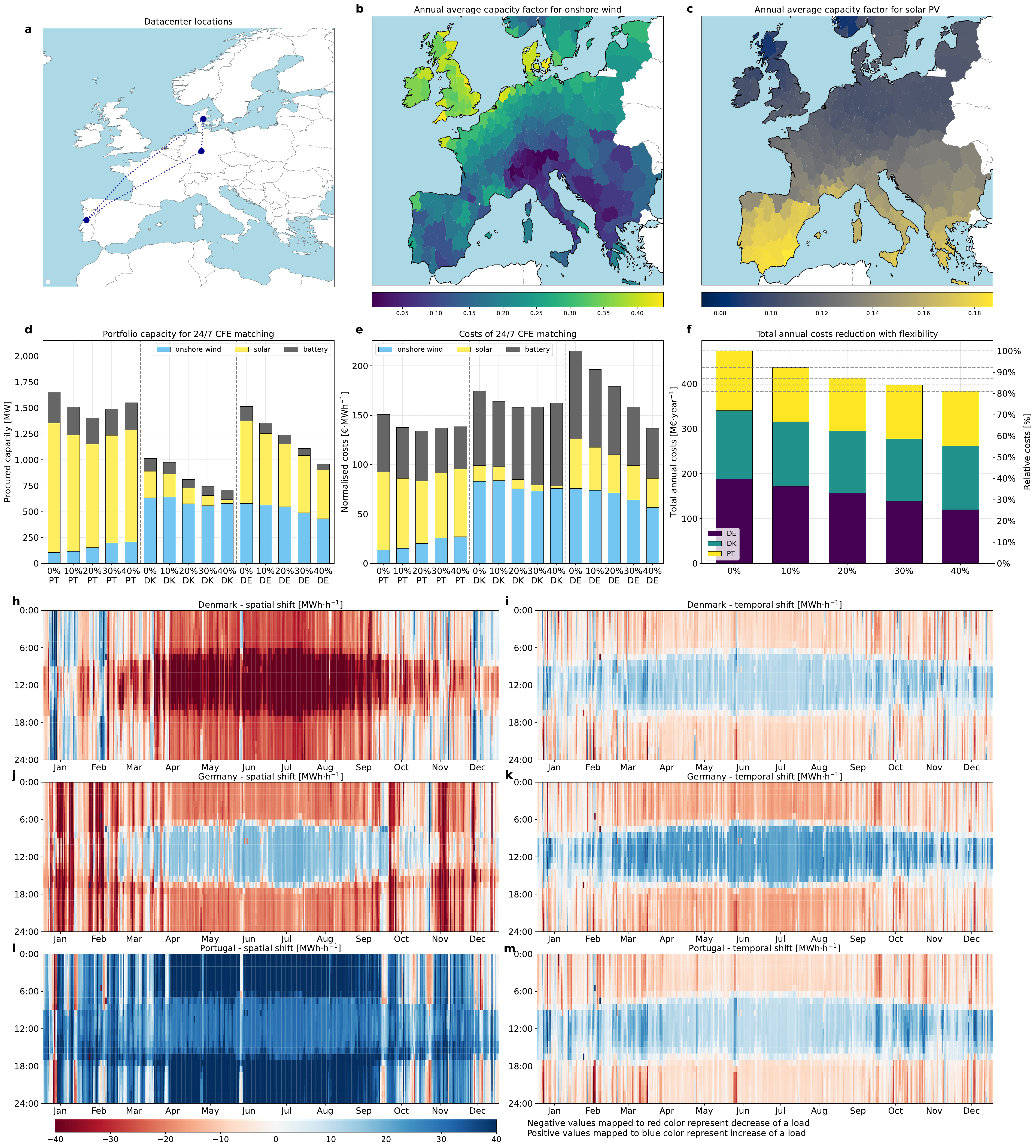}
    \caption{Illustration of the signal 1: quality of local renewable resources.
        \textbf{a:} Assumed datacenter locations: Denmark, Germany, Portugal.
        \textbf{b,c:} Annual average capacity factor of onshore wind and solar photovoltaic. Data is simulated using the ERA5 reanalysis dataset for weather year 2013 and aggregated to 256 regions in Europe.
        \textbf{d:} Cost-optimal portfolio of renewable resources and battery storage sufficient for 24/7 matching. Steps on x-axis represent increasing share of flexible load.
        \textbf{e:} Cost breakdown of 24/7 matching strategy.
        \textbf{f:} Total annual costs of 24/7 matching strategy as a function load flexibility. Relative axis is normalized to the costs of inflexible load.
        \textbf{h,j,l:} Hourly spatial load shifts for the three datacenter locations. Color mapping represents the quantity of load \enquote{received} from other locations (positive values) or \enquote{sent} away (negative values).
        \textbf{i,k,m:} Hourly temporal load shifts for the three datacenter locations. Color mapping represents the quantity of load shifted to a given hour from other times, or shifted from a given hour to another time.}
    \label{fig:dashboard1}
\end{figure*}

First, we focus on the signal associated with the difference in quality of local renewable resources. Consider a case when datacenters are located in Denmark, Germany, and Portugal (Figure \ref{fig:dashboard1}: panel a). There are three different types of renewable resources represented by the three sites, with Denmark having a good wind resource, Germany having a moderate wind and solar resources, and Portugal having a good solar resource. The quality of local renewable resources is reflected by the average annual capacity factor for onshore wind and solar PV, as shown in the panels b and c.

Figure \ref{fig:dashboard1}: panel d shows the modelled cost-optimal capacity of renewable generators and battery storage for 24/7 matching. In the case of inflexible loads, the procurement strategy must ensure a sufficient amount of carbon-free electricity is available round-the-clock \textit{at each site}. Due to the non-dispatchable nature of renewable resources and the relatively high energy costs of battery storage, the cost-optimal portfolio is much larger than the load. To cover the 100 MW load with perfect hourly matching, the consumer would have to procure combined wind and solar PV generators of approx. 900~MW in Denmark and 1300~MW in Germany or Portugal. In each portfolio, the proportion of wind and solar PV reflects the quality of the local resources.

In Figure \ref{fig:dashboard1}: panel d, steps on the x-axis represent increasing flexible load, allowing for a reduction in the capacity of renewable generators and battery storage necessary for 24/7 matching. By having a degree of load flexibility, a consumer can shift the load from times of scarce renewable generation to times of abundant renewable generation in the same hour, or use the excess renewable generation at one site to cover the load at another site.

The breakdown of costs associated with each procurement strategy is shown in Figure \ref{fig:dashboard1}: panel e. The panel shows the average cost per MWh of consumption, including renewable generation costs and battery storage costs.\footnote{Here we do not consider the option of selling the excess electricity to the regional grid. Excess electricity can thus be stored in the battery, shifted in space or time with load flexibility, or curtailed.} Demand flexibility enables consumers to better match renewable generation with demand profile, reducing over-procurement and costly battery storage. The costs are reduced in all locations, and especially in locations where hourly matching is most expensive. Modelling results show the costs of 24/7 matching in Germany are at 215~\euro/MWh when there is no load flexibility, 195~\euro/MWh when there is 10\% flexible loads, and reduced up to 137~\euro/MWh when there is 40\% of flexible loads.

Figure \ref{fig:dashboard1}: panel f displays the total annual costs for achieving 24/7 matching at all locations (left y-axis) and their relative representation as a percentage of the inflexible scenario costs (right y-axis). This panel summarises the model results: as demand flexibility increases, 24/7 matching becomes more resoure-efficient and affordable. The value of additional flexibility also diminishes as the number of flexible loads increases, i.e., the first 10\% of flexible loads can reduce costs more than the next 10\%. A generalization of this result is presented in Section \ref{ssec:section4}.

Finally, Figure \ref{fig:dashboard1}: panels h-m show optimal hourly load shifts for the three datacenter locations based on a selected scenario with 40\% of flexible loads. Several observations can be drawn from the plots regarding the use of spatial and temporal flexibilities.

First, there are clear daily and seasonal patterns of spatial load shifts (panels h, j, l) driven by the quality of local resources, such as wind or solar PV capacity factors. For renewable generators, better resource quality implies higher energy yield per MW of installed capacity, which translates into lower Levelised Costs of Electricity (\gls{lcoe}). Whenever spatial load shifting is possible, a rational strategy is \textit{to get the most out of the clean energy resources with good quality}. The heatmaps above illustrate this behavior well: a datacenter located in Denmark---a region with poor solar resources---tends to shift loads away from the mid-spring till mid-autumn. Instead, a datacenter located in Portugal---a region with better solar resources---tends to receive loads during this period. It works just about reciprocally for wind-related load shifts: a datacenter in Portugal benefits from having a partner in Denmark, the very windy region in Europe.
Germany's datacenter is located between the two extremes, receiving workloads from Denmark during summer daytime and moving them to other locations at other times.

Second, temporal load shifts (panels i, k, m) is used to \enquote{flatten} the renewable generation profile. The heatmaps show that the load is consistently shifted to the daytime hours, when solar generation is abundant, and away from the nighttime hours. Note that the temporal flexibility is subject to the daily usage conservation constraint (eq. \ref{eqn:dailyconserv2}), which limits the temporal flexibility usage for longer periods.

\subsection{Signal 2: low correlation of wind power generation over long distances}

\begin{figure*}[h]
    \centering
    \includegraphics[width=1\textwidth]{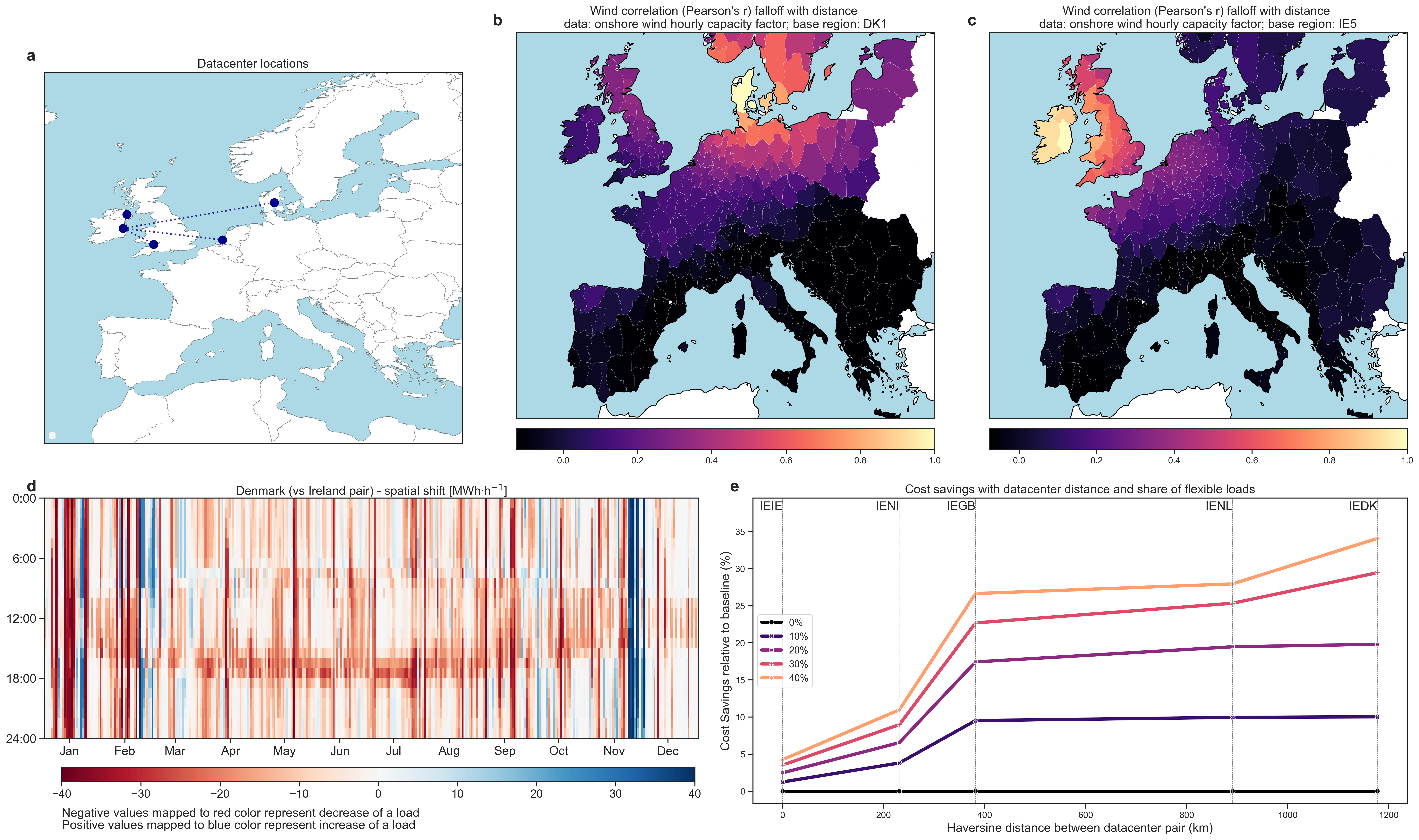}
    \caption{Illustration of the signal 2: low correlation of wind power generation over long distances.
    \textbf{a:} Assumed datacenter locations: pairwise connections across regions with similar quality of renewable resources. Here: Ireland with Northern Ireland, or England, or the Netherlands, or Denmark (-west zone).
    \textbf{b,c:} Peason correlation of hourly capacity factor for onshore wind generation, if Denmark (-west zone, panel b) or selected region of Ireland (panel c) are taken as basis. As a result of different weather conditions, wind feed-in has a noticeable correlation falloff over distances of 300-400 km. Data is simulated using the ERA5 reanalysis dataset for weather year 2013 and aggregated to 256 regions in Europe.
    \textbf{d:} Hourly spatial load shifts for the selected scenario and datacenter; here datacenter is located Denmark (and another one is located Ireland). Color mapping represents the quantity of load \enquote{received} from other locations (positive values) or \enquote{sent} away (negative values).
    \textbf{e:} Cost savings of 24/7 matching with increasing distance between datacenters. Costs are normalized to the cost level of inflexible load.}
    \label{fig:dashboard2}
\end{figure*}

Second, we focus on the signal associated with low correlation over long distances in wind power generation. Suppose a pair of datacenters are located so that their local wind resources are high quality, but their solar resources are low quality, with increasing distance between them. Taking five pairs as an example, we can consider Ireland and Northern Ireland, England, the Netherlands, and Denmark (Figure \ref{fig:dashboard2}: panel a).

We can use historical weather data to calculate the Pearson correlation coefficient of the  hourly capacity factors for onshore wind generation. Figure \ref{fig:dashboard1}: panels b and c display the result for the selected regions in Ireland and Denmark, accordingly. There is a drastic falloff in wind feed-in correlation over distances of 300-400~km due to varying weather conditions. This calculation replicates observations in a previous study on the statistical properties of geographically dispersed wind power \cite{hascheGeneralStatisticsGeographically2010}. It follows that the feed-in of wind power generators procured, for instance, by datacenters in Ireland and Denmark is not correlated.

By using spatial flexibility, datacenter operators can do "load arbitrage" between locations with different weather conditions to take advantage of wind generation's stochastic nature. In the heatmap of spatial load shifts for selected datacenter pair (Figure \ref{fig:dashboard1}: panel d), the geographical load shifts resulting from uncorrelated wind feed-in can be spotted by vertical stripes (sudden change of direction) with stochastic patterns. Note that there is no strong seasonal pattern, as it was observed in the previous example, since both locations share similar quality of wind and solar resources.

Figure \ref{fig:dashboard1}: panel e illustrates the relationship between the distance between datacenter pairs and the resultant cost savings. The results show that the cost savings are most noticeable for steps of 300-400~km, which is consistent with the falloff in wind generation correlation. As the load is more flexible, more load arbitrage is possible, resulting in cost savings; if the load is inflexible, the datacenter pair cannot harvest the benefits of different local weather conditions. In distances greater than 300-400~km, wind feed-in correlations are already low, which leads to no significant cost savings. Distances over 1000~km and high flexibility levels result in a slight increase in cost savings due to the time lag between solar generation peak times.  The effect is visible on the heatmap of spatial load shifts. We discuss this signal in detail in the next section.

\subsection{Signal 3: time lag in solar radiation peaks due to Earth's rotation}

\begin{figure*}
    \centering
    \includegraphics[width=\textwidth]{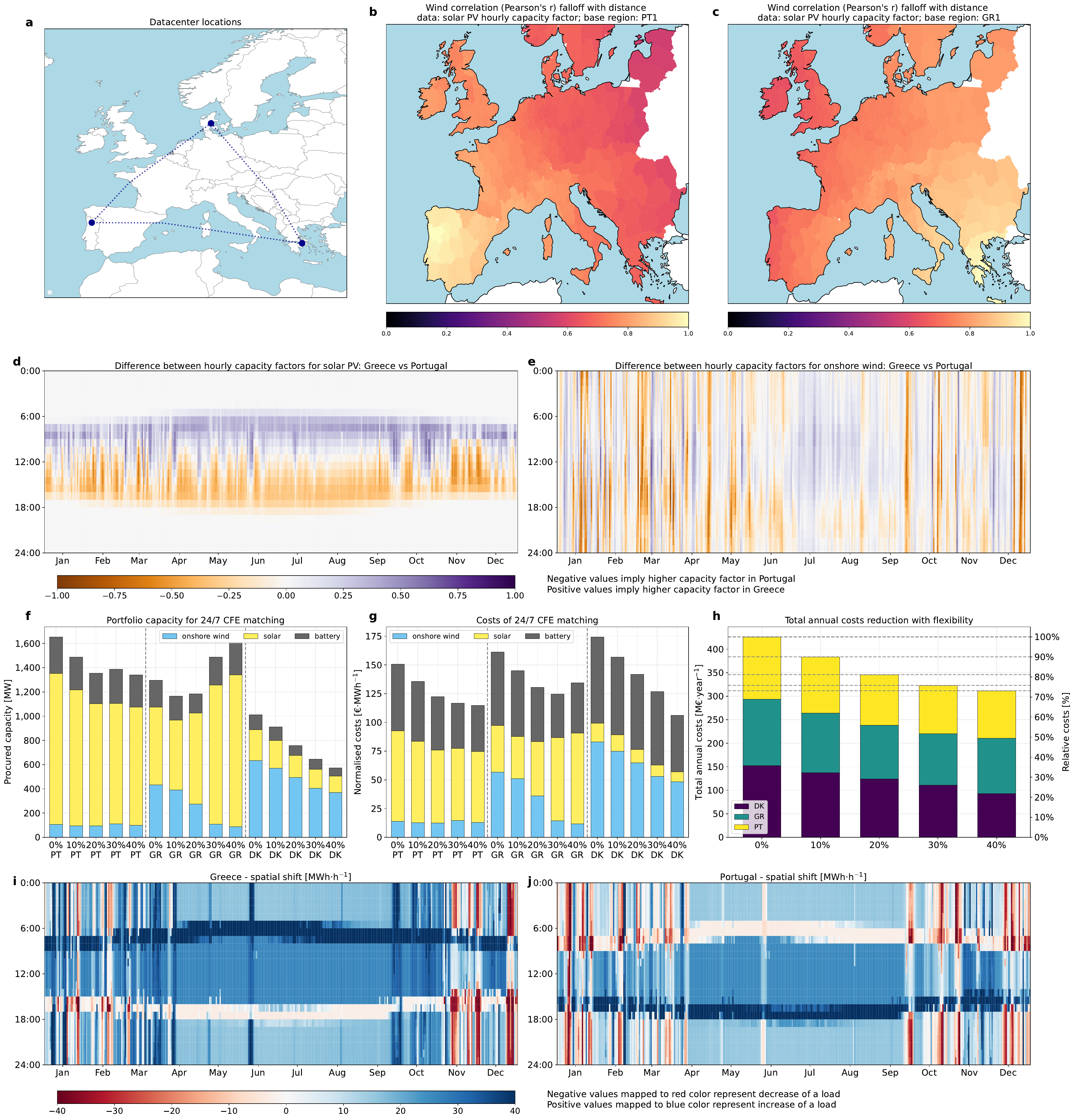}
    \caption{Illustration of the signal 3: time lag in solar radiation peaks due to Earth's rotation.
    \textbf{a:} Assumed datacenter locations: Denmark, Portugal, Greece.
    \textbf{b,c:} Peason correlation of hourly capacity factor for solar photovoltaic generation, if selected region of Portugal (panel b) or region of Greece (panel c) is taken as basis. Solar generation remains highly correlated over long distances, in contrast to wind generation.
    \textbf{d:} Difference in solar photovoltaic hourly capacity factors between two selected locations: Greece and Portugal. The two locations are approx. 2700~km apart, which results in a noticeable lag in solar generation peaks due to Earth's rotation.
    \textbf{e:} Difference in wind generation hourly capacity factors between two selected locations: Greece and Portugal. As expected, low correlation of wind feed-in over long distance results in stochastic pattern.
    \textbf{f:} Cost-optimal portfolio of renewable resources and battery storage sufficient for 24/7 matching. Steps on x-axis represent increasing share of flexible load.
    \textbf{g:} Cost breakdown of 24/7 matching strategy.
    \textbf{h:} Total annual costs of 24/7 matching strategy as a function load flexibility. Relative axis is normalized to the costs of inflexible load.
    \textbf{i,j:} Hourly spatial load shifts for the selected datacenter locations: Greece (panel i) and Portugal (panel j). Color mapping represents the quantity of load \enquote{received} from other locations (positive values) or \enquote{sent} away (negative values).}
    \label{fig:dashboard3}
\end{figure*}

Third, we focus on the signal associated with the time lag in solar radiation peaks due to Earth's rotation. Consider a case when datacenters are located in Denmark, Greece, and Portugal (Figure \ref{fig:dashboard3}: panel a). In this scenario, Danish datacenter has access to high-quality wind resources, whereas Greek and Portuguese datacenters have access to good solar resources.

The Pearson correlation of the simulated hourly capacity factors for solar photovoltaic generation (Figure \ref{fig:dashboard3}: panels b and c) show that solar generation has good correlation over long distances, in contrast to wind. However, the datacenters in Greece and Portugal are approx. 2700~km apart, which creates potential for the use of spatial flexibility to take advantage of the time lag in solar generation peaks. The difference in solar photovoltaic hourly capacity factors between the two selected locations is shown in Figure \ref{fig:dashboard3}: panel d. The panel illustrates the time lag in solar generation peaks due to Earth's rotation. The peak of solar generation in Portugal is delayed by several hours compared to Greece. For comparison, the difference in wind generation hourly capacity factors between the two selected locations is shown in Figure \ref{fig:dashboard3}: panel e. As expected, low correlation of wind feed-in over long distance results in a stochastic pattern.

Panels f, g and h in Figure \ref{fig:dashboard3} show the cost-optimal portfolio of renewable resources and battery storage for 24/7 matching, cost breakdown, and total annual costs of 24/7 matching strategy as a function of load flexibility. Similar to the previous examples, the cost-optimal portfolio of resources necessary for 24/7 matching and associated costs are reduced with increasing flexible loads, as datacenters can get the most out of the clean energy resources with good quality, as well as take advantage of the time lag in solar generation peaks. In the case of the Danish datacenter, benefiting from having partners in solar-rich locations, the costs of 24/7 matching strategy is reduced from 174~\euro/MWh to 157~\euro/MWh with 10\% of flexible loads, and down to 106~\euro/MWh with 40\% of flexible loads.

Finally, Figure \ref{fig:dashboard3}: panels i and j show optimal hourly load shifts for the selected datacenter locations in Greece and Portugal, accordingly. The heatmaps illustrate how spatial flexibility can be used to benefit from the time lag in solar generation peaks. The datacenters in Greece and Portugal receive loads from Denmark between midspring and midfall, with a clear daily pattern caused by the time lag in solar generation peaks. Additionally, the heatmaps show that during the winter months, when solar generation is scarce, both datacenters tend to shift loads to Denmark.

\subsection{Generalising the results beyond specific load locations}
\label{ssec:section4}

Finally, we generalise the results beyond specific load locations and to provide a more comprehensive understanding of the relationship between load flexibility and the costs of 24/7 CFE matching.

Consider that datacenters can be located in any of the following eight countries: Denmark, Ireland, the Netherlands, Germany, Latvia, Greece, Portugal, and France. We selected these countries to represent a variety of renewable resources and geographical locations.\footnote{The analysis could be expanded to include additional locations, but the number of combinations would be significantly increased. These locations are a reasonable compromise between generalisation quality and computational demands.} The analysis is carried out by modeling all combinations of three datacenters in the selected countries. With three countries selected from eight, and the order of the countries being irrelevant, there are 56 possible combinations. As a result, the scenario space includes combinations of datacenters with different quality of renewable resources, as well as different distances between them.
For each combination of datacenters and level of flexibility, we conduct the same analysis as in the previous sections, i.e., the model solution includes information about the cost-optimal portfolio of renewable resources and battery storage for 24/7 matching, procurement policy cost breakdowns, and optimal utilisation of spatial and temporal load flexibility.

\begin{figure}
    \centering
    \includegraphics[width=1\columnwidth]{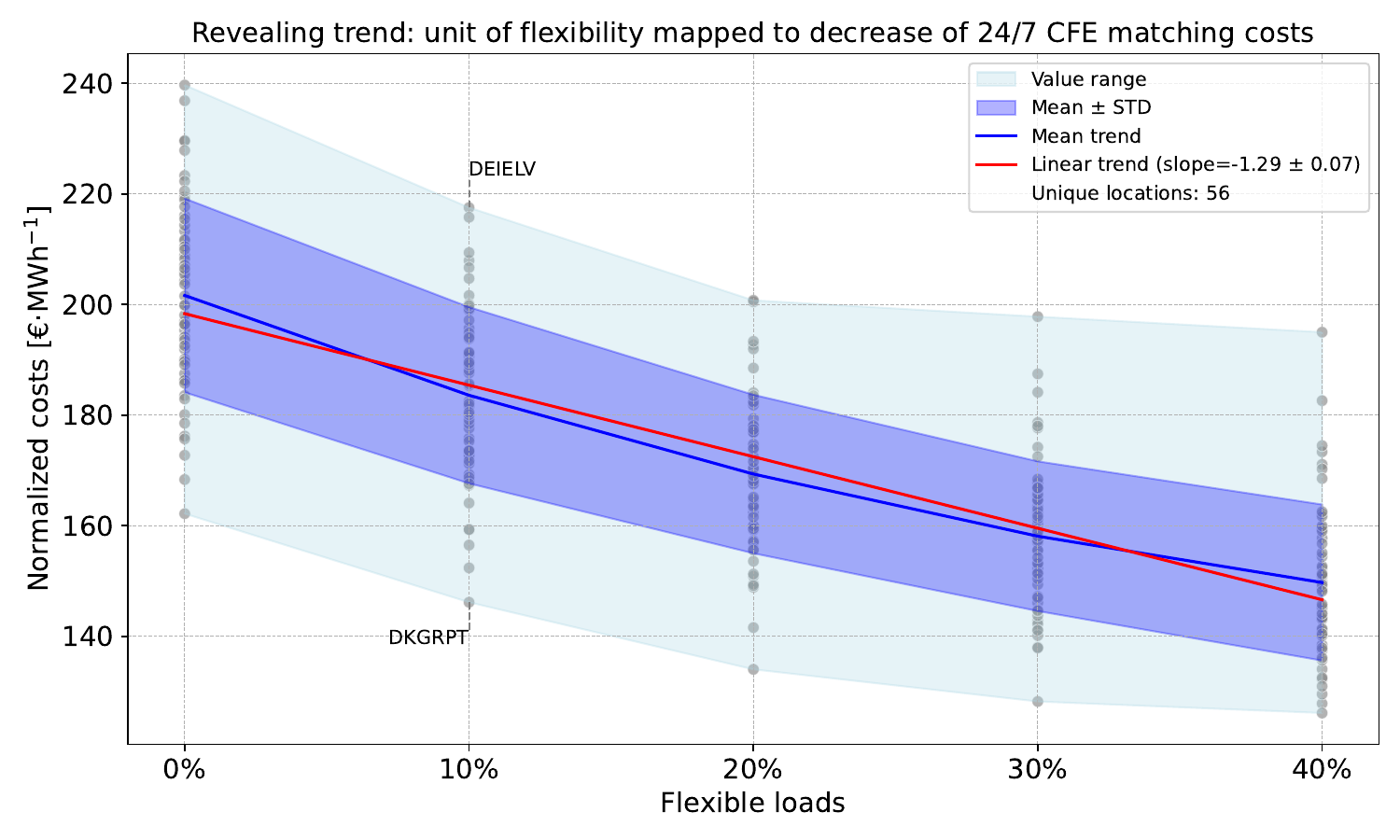}
    \caption{Relationship between the level of flexibility and the costs of 24/7 matching for 56 combinations of datacenter locations. The costs are normalised to the cost level of inflexible load.}
    \label{fig:flexcost}
\end{figure}

The results of the analysis are presented in Figure \ref{fig:flexcost}. The figure shows the relationship between the level of flexibility and the costs of 24/7 matching for all 56 combinations of datacenters. The normalised costs of 24/7 matching lie in a broad range from 240~\euro/MWh to ca. 160~\euro/MWh for inflexible load scenario. The lower bound represents combination Denmark-Greece-Portugal, since these locations share excellent quality of wind and solar resources. The upper bound represents a combination of Germany, Ireland, and Latvia, since none of the three locations have a good solar resource, and only Ireland has a good wind resource.  As a result, 24/7 CFE matching is especially expensive in two out of three locations -- Germany and Latvia.

Several conclusions can be drawn from the analysis. First, space-time load-shifting flexibility facilitates the efficiency and affordability of 24/7 CFE matching, regardless of the location of the datacenters. Based on specific datacenter locations, the optimal use of space-time flexibility may be driven by one, or a combination, of the signals discussed above; however, in general, flexibility is beneficial to all locations.

Second, the costs of 24/7 CFE matching are reduced by 1.29$\pm$0.07~\euro/MWh for every additional percentage of flexible load. The low standard error suggests that the relationship is robust across different combinations of datacenters. This finding implies that datacenters can make use of the available flexibility to reduce the costs of 24/7 CFE matching, regardless of which of the three signals is dominant for the specific locations. For example, if renewable energy quality is similar across all locations, and solar PV has a high \gls{lcoe}, the two related signals have little impact on optimal space-time shifting and 24/7 CFE procurement strategy. However, when this occurs, the company operating the datacenters can use all the available load flexibility to \enquote{arbitrage} on low correlation of wind feed-in, and factor this into the procurement strategy.

Third, the overlap between the mean trend and linear regression line suggests that the value of additional flexibility diminishes as the number of flexible loads increases, i.e., the first 10\% of flexible loads can reduce costs more than the next 10\% of flexible loads. The latter is consistent with the observations made in previous sections.

\section{Discussion}
\label{sec:discussion}
%

\textbf{Our results in perspective --} This work contributes to and builds on two streams of literature: one is concerned with the flexibility and environmental impact of datacenters, the another with the means, costs and implications of 24/7 carbon-free energy matching.

Regarding the first literature stream, our findings are in line with those of existing model-based studies on the flexibility potentials offered by geographically distributed datacenters: load-shifting flexibility facilitates the integration of renewable energy sources and reduces the carbon footprint of datacenters \cite{zhengMitigatingCurtailmentCarbon2020, mahmudDistributedFrameworkCarbon2016, wangGreenawareVirtualMachine2015, kimDataCentersDispatchable2017, liuGeographicalLoadBalancing2011, kellyBalancingPowerSystems2016, lindbergEnvironmentalPotentialHyperScale2021}.
Although opportunities and challenges of datacenter flexibility have drawn considerable attention in the literature, space-time load-shifting flexibility has not been studied in the context of 24/7 CFE matching pursued by companies aiming to run their facilities at 0~gCO$_2$/kWh.
Since hourly matching of consumption with carbon-free electricity requires and implies renewable energy procurement through \gls{ppa}s and thus less electricity procurement from the grid, we argue that grid signals such as average carbon emission intensity or locational electricity prices have low value for informed use of load-shifting flexibility in the context of the 24/7 CFE matching.
In order to address this challenge, we identify three signals that can enable companies to shift load across space and time effectively.
Moreover, we demonstrate how companies can achieve significant gains in energy efficiency and affordability of 24/7 CFE matching by including these signals into their energy procurement and load-shaping strategies.

Regarding the second literature stream, our results indicate that space-time load-shifting increases access to clean electricity and provides flexibility to consumers for matching demand with carbon-free electricity.
As a result, datacenters and other commercial and industrial consumers who have flexible demands can achieve high degrees of carbon-free energy matching with lower resources and thus be more cost-effective.
The system decarbonization impact of 24/7 CFE procurement indicated by the model-based studies \cite{xu-247CFE-report,riepin-zenodo-systemlevel247} could therefore be amplified with greater participation.

Finally, our results follow the general consensus on the benefits of geographical load balancing in highly renewable electricity networks \cite{schlachtbergerBenefitsCooperationHighly2017}.
However, the benefits of connecting dispersed regions at continental scales are always weighed against the high costs and long lead times associated with grid expansion and reinforcement.
It is unique to the datacenter industry that geographical load-shifting flexibility can be achieved at low costs and over long distances by harnessing local differences in the quality of renewable resources and renewable generation profiles.

\textbf{A broader context --} The first commitments to 24/7 CFE matching initiated by the datacenter industry can help companies from other sectors by establishing procurement methods and standard practices for 24/7 CFE matching. \cite{xu-247CFE-report}.
In the same way, early efforts to consider potentials of available load flexibility within energy procurement and load-shifting strategies can help a broad range of companies that are seeking to lower their carbon footprints.
In the event that datacenter companies, as well as other commercial and industrial consumers with flexible loads can achieve high degrees of 24/7 CFE matching at moderate costs, more actors may be willing to commit to it.
Therefore, the system decarbonization impacts associated with the 24/7 CFE procurement \cite{riepinMeansCostsSystemlevel2023} could be amplified with greater participation while requiring fewer resources.

Additionally, even though it is not the focus of the present study, our results provide necessary methodological foundation for studying strategic locations for new datacenters by companies interested to reduce or eliminate completely their carbon footprint. It is therefore important to consider the factors facilitating resource-efficient and cost-effective 24/7 CFE matching identified by this study when choosing optimal locations for datacenters.

\textbf{Critical appraisal and further work --} The study design includes simplifications and assumptions inherent to energy optimization models. The results should thus be seen with caution, i.e., as model-derived insights rather than quantitative projections.

One of the central assumptions of the model experiment is that datacenters achieve a certain degree of flexibility in their workloads. While we control the share of flexible workloads exogenously (see \nameref{sec:methods}), the actual flexibility of datacenters is not known with certainty and can vary significantly among facilities.
As specific data becomes available, the model can be calibrated to reflect the actual flexibility potential.

Additional empirical research is required to quantify the costs and benefits of utilizing demand flexibility in the \gls{ict} industry.
Further studies could usefully explore the costs and technical potentials of achieving a certain share of flexible workloads, which are not considered in this study.
By including implicit flexibility costs, the benefits of flexibility can be quantified more accurately.
An empirical improvement could also address technical aspects and properties of flexible workloads, such as power usage ramping up and down, reliability, and performance constraints.

In the context of 24/7 CFE matching, further research is needed to capture the interaction of promising energy technologies, such as long-term duration energy storage and space-time load-shifting flexibility.
This case is particularly relevant, since 24/7 CFE procurement could create an early market and drive deployment of advanced technologies \cite{xu-247CFE-report,riepinMeansCostsSystemlevel2023}.
It was also demonstrated that geographical load shifting complements long-term duration storage \cite{riepinValueSpacetimeLoadshifting2023}.

\section{Conclusion}
\label{sec:conclusion}
%

This work explores the role of space-time load-shifting flexibility provided by geographically distributed datacenters in achieving 24/7 carbon-free energy (CFE) matching and identifies signals relevant for effective load shifting.

We develop an optimization model to simulate a network of datacenters managed collectively by a company pursuing 24/7 carbon-free energy matching objective.
The model provides users with the ability to specify datacenter locations within the European electricity system, along with the flexibility potentials, among other parameters.
Through energy system modeling, we isolate three individual signals that are crucial for effective space-time load-shifting: (i) varying average quality of renewable energy resources across datacenter locations, (ii) low correlation between wind power generation over long distances due to different weather conditions, and (iii) lags in solar radiation peak due to Earth's rotation.

In relation to the first signal, we demonstrate that leveraging spatio-temporal load flexibility maximizes the utility of high-quality renewable energy resources. Therefore, the optimal load-shaping strategy entails making informed decisions about where and when to shift loads, based on the average capacity factors of procured renewable energy resources at different locations. Our results indicate that energy costs are reduced at all locations, particularly in those where hourly matching is most challenging. For instance, for a data center located in Germany with partners in Denmark and Portugal, the costs of 24/7 CFE matching decrease from 215~\euro/MWh with no load flexibility to 195~\euro/MWh with 10\% flexible loads, and further to 137~\euro/MWh with 40\% flexible loads.

As for the second signal, we show that spatio-temporal load shifting allows datacenter companies to take advantage of the low correlation between wind power generation over long distances, i.e., to perform \enquote{load arbitrage} between locations with different weather conditions.
The results show that the cost savings are most noticeable for steps of 300-400 km, which is in line with the falloff in wind generation correlation.
Considering a pair of datacenters located approximately 380~km apart in regions with similar wind and solar conditions, the introduction of 10\% flexible loads yields a 9.5\% reduction in energy costs, whereas 40\% flexible loads facilitate a 26.5\% cost savings.
For distances greater than 400~km, wind feed-in correlations are already low, so further increasing the distance between datacenters does not lead to additional cost savings.

Regarding the third signal, we show how spatio-temporal load shifting can be used to exploit the lags in solar radiation peak due to Earth's rotation. Thus, companies seeking 24/7 CFE matching can benefit from the time zone differences between datacenters and the hourly capacity factors of solar power generators at different locations. Solar power can be used more efficiently this way, thereby reducing costs associated with carbon-free energy matching.

Finally, we generalize our results to a broader set of datacenter locations across Europe and show the relationship between the load flexibility and the costs of 24/7 CFE matching. The results reveal that the resource-efficiency and cost-effectiveness of hourly matching are improved by the effective use of space-time load flexibility in all cases. However, the location of datacenters and the time of year affect which of the three signals are most relevant for an effective load-shaping strategy. Further, we show that with every percentage of flexible load added, the costs of 24/7 CFE matching fall on average by 1.29$\pm$0.07 \euro/MWh, which can be interpreted as the marginal value of load flexibility. Our results also illustrate the diminishing returns of additional load flexibility, which is important for companies seeking to co-optimise their long-term energy procurement and short-term load-shifting strategies.

Overall, our results indicate that space-time load-shifting flexibility increases access to clean electricity and gives consumers more options to match demand with carbon-free electricity. Demand flexibility can enhance resource efficiency and reduce costs of truly clean computing. This can amplify the decarbonization effects associated with the voluntary clean energy commitments by encouraging more datacenter companies and other electricity consumers to join the 24/7 CFE movement.

\section*{Acknowledgements}

We thank the following people for support and fruitful discussions on the art of energy system modelling: Elisabeth Zeyen, Fabian Hofmann, Fabian Neumann, Devon Swezey, Ana Radovanovic, Chris Adams, Kristina Riepin and the research staff of the Scalable Systems Laboratory at the University of Wisconsin-Madison.

\section*{Author Contributions}


\textbf{Iegor Riepin}:
Conceptualization --
Data curation --
Formal Analysis --
Investigation --
Methodology --
Software --
Validation --
Visualization --
Writing - original draft
\textbf{Tom Brown}:
Conceptualization --
Methodology --
Formal Analysis --
Software --
Supervision --
Writing - review \& editing
\textbf{Victor M. Zavala}:
Conceptualization --
Methodology --
Formal Analysis --
Writing - review \& editing

\section*{Data and Code Availability}
\label{sec:code}

The simulations were carried out with PyPSA --- an open-source software framework for simulating and optimising modern energy systems \cite{brownPyPSAPythonPower2018}.
All code, input data and results are published under an open license. The code to reproduce the experiments is available at GitHub \cite{github-spacetime}.

\printglossary[type=\acronymtype]

\addcontentsline{toc}{section}{References}
\renewcommand{\ttdefault}{\sfdefault}
\bibliography{bibliography}


\newpage

\makeatletter
\renewcommand \thesection{S\@arabic\c@section}
\renewcommand\thetable{S\@arabic\c@table}
\renewcommand \thefigure{S\@arabic\c@figure}
\makeatother
\renewcommand{\citenumfont}[1]{S#1}
\setcounter{equation}{0}
\setcounter{figure}{0}
\setcounter{table}{0}
\setcounter{section}{0}

\section*{Supplementary Information}
\label{sec:si}

\section{Technology assumptions}
\label{sec:si_1}
%


Cost and other assumptions for energy technologies available for 24/7 CFE participating consumers are collected from the Danish Energy Agency \cite{DEA-technologydata}. These assumptions are listed in \cref{tab:tech_costs}.
A full list of technology assumptions, including the data for energy technologies in the background energy system, is available via the reproducible scientific workflow in the GitHub repository \cite{github-spacetime}.

\begin{table*}[t]
    \centering
    \resizebox{0.95\textwidth}{!}{%
        \begin{tabular}{lccccccc}
            \hline\hline
            \textbf{Technology} &
            \textbf{Year} &
            \textbf{\begin{tabular}[c]{@{}c@{}}CAPEX\\ (overnight cost)\end{tabular}} &
            \textbf{\begin{tabular}[c]{@{}c@{}}FOM\\ {[}\%/year{]}\end{tabular}} &
            \textbf{\begin{tabular}[c]{@{}c@{}}VOM\\ {[}Eur/MWh{]}\end{tabular}} &
            \textbf{\begin{tabular}[c]{@{}c@{}}Efficiency\\ {[}per unit{]}\end{tabular}} &
            \textbf{\begin{tabular}[c]{@{}c@{}}Lifetime\\ {[}years{]}\end{tabular}} &
            \textbf{Source} \\ \hline\hline
            Utility solar PV & 2025 & 612 \officialeuro/kW & 1.7 & 0.01 & - & 37.5 & \cite{DEA-technologydata} \\
            Onshore wind & 2025 & 1077 \officialeuro/kW & 1.2 & 0.015 & - & 28.5 & \cite{DEA-technologydata} \\
            Battery storage & 2025 & 187 \officialeuro/kWh & - & - & - & 22.5 & \cite{DEA-technologydata} \\
            Battery inverter & 2025 & 215 \officialeuro/kW & 0.3 & - & 0.96 & 10 & \cite{DEA-technologydata} \\
            \hline \hline
        \end{tabular}%
    }
\begin{tablenotes}
    {\footnotesize
    \item[] Notes: All costs are in 2020 euros; CAPEX = capital expenditure; FOM = fixed operations and maintenance costs; VOM = variable operations and maintenance costs.
    }
\end{tablenotes}
    \vspace{0.2cm}
    \caption{Technology assumptions.}
    \label{tab:tech_costs}
\end{table*}


\end{document}